\documentstyle[12pt,psfig]{article}
\makeatletter
\def\eqalign#1{\null\vcenter{\def\\{\cr}\openup\jot\m@th
  \ialign{\strut$\displaystyle{##}$\hfil&$\displaystyle{{}##}$\hfil
      \crcr#1\crcr}}\,}
\makeatother
\def\mod{{\rm mod}}
\begin{document}
\bigskip
\bigskip
\bigskip
\begin{center}
{\Large\bf 
Bethe ansatz for the Harper equation: Solution for a small commensurability
parameter. 
}\\
\bigskip
\bigskip
\bigskip
\bigskip
{\large I. V. Krasovsky}\\
\bigskip
Max-Planck-Institut f\"ur Physik komplexer Systeme\\
N\"othnitzer Str. 38, D-01187, Dresden, Germany\\
E-mail: ivk@mpipks-dresden.mpg.de\\
(present address)\\
\medskip
and\\
\medskip
B.I.Verkin Institute for Low Temperature Physics and Engineering\\
47 Lenina Ave., Kharkov 310164, Ukraine.
\end{center}
\bigskip
\bigskip
\bigskip

\noindent
{\bf Abstract.}
The Harper equation describes an electron on a 2D lattice in magnetic field
and a particle on a 1D lattice in a periodic potential, in general,
incommensurate with the lattice potential.
We find the distribution of the roots of Bethe ansatz equations
associated with the Harper equation
in the limit as $\alpha=1/Q\to 0$, where $\alpha$ is the commensurability 
parameter ($Q$ is integer).
Using the knowledge of this distribution we calculate the higher and 
lower boundaries of the spectrum of the 
Harper equation for small $\alpha$. The result is in agreement with the
semiclassical argument, which can be used for small $\alpha$.

\newpage
\section{Introduction}
The Harper equation (see [\ref{Har}-\ref{AA}] and also
[\ref{HK},\ref{Last1},\ref{KS}] for recent reviews)
\begin{equation}
\psi_{n-1}+2\cos(2\pi\alpha n +\theta)\psi_n
+\psi_{n+1}=\varepsilon\psi_n,\qquad
n=\dots,-1,0,1,\dots,\qquad \alpha,\theta\in{\bf R}\label{Harper}
\end{equation}
in $l^2(Z)$ describes a quantum particle on the line in two periodic
potentials which are incommensurate if $\alpha$ is irrational. It is also 
a model for 
an electron on a square lattice subject to a perpendicular
uniform magnetic field (Azbel-Hofstadter). In the latter case $\alpha$ is
proportional to the value of the magnetic field and $\theta$ is a 
quasimomentum
in the direction along one of the sides of the elementary square  
(see, e.g, [\ref{WZ}]). The Harper equation has many deep connections to
various fields of modern physics, e.g., quasicrystals, quantum Hall effect,
and pure mathematics: number theory, functional analysis.

Equation (\ref{Harper}) is the eigenvalue equation for a particular case 
of the almost Mathieu operator. 
The spectrum $\sigma_M(\alpha,\theta)$ of the corresponding operator
consists of $Q$ intervals (bands) if $\alpha=P/Q$,
where $P$ and $Q$ are coprime integers\footnote{That is they do not have 
a common divisor other than 1.}.
The spectrum $\sigma_H(\alpha)$ of the Azbel-Hofstadter model is the union 
over all real $\theta$ of $\sigma_M(\alpha,\theta)$. At $\alpha=P/Q$ the
spectrum  $\sigma_H(P/Q)$ consists of $Q$ bands.
If $\alpha$ is irrational, it is known that $\sigma_M(\alpha,\theta)$ is 
independent of $\theta$, and hence the spectrum
$\sigma_H(\alpha)=\sigma_M(\alpha,\theta)$.
The plot of $\sigma_H(\alpha)$ as a function of $\alpha$ is
called the Hofstadter butterfly [\ref{Hof}].
The case of irrational $\alpha$ is the most interesting.
There is a conjecture that the measure of the spectrum is zero for
all irrational $\alpha$ [\ref{Last2},\ref{AMS},\ref{T}].
This and many other questions regarding the spectrum of the almost 
Mathieu operator at irrational $\alpha$ are still unresolved 
(see [\ref{Last1},\ref{WZ},\ref{ATW}] for a list of problems). 

Recently a new approach to these problems has been proposed 
[\ref{WZ},\ref{WZMPL}].
Let $\alpha=P/Q$.
First, it can be shown [\ref{WZ}] that certain $Q$ (one from each band) 
points of the spectrum\footnote{
This is true both for $\sigma_M(P/Q,\theta)$ and for $\sigma_H(P/Q)$.}
are the solutions $\varepsilon$ of the 
following eigenvalue equation:
\begin{equation}
i(z^{-1}+qz)\Psi(qz)-i(z^{-1}+q^{-1}z)\Psi(q^{-1}z)=\varepsilon\Psi(z)
\label{de}
\end{equation}
in the space of polynomials $\Psi(z)=\prod_{k=1}^{Q-1}(z-z_k)$ of degree $Q-1$.
Here $q=e^{i\gamma}$, $\gamma=\pi\alpha=\pi P/Q$.
This equation is related to the algebra $U_q(sl_2)$.
One can obtain other difference equations similar to (\ref{de}) with the same
set of eigenvalues. Analogous equations are also available for other 
points of the spectrum $\sigma_M(P/Q,\theta)$
[\ref{ATW},\ref{FK}]. In the Appendix we present 
a simple derivation of (\ref{de}) which is based on the fact that it
can be regarded as an equation for the generating function of a finite system 
of orthogonal polynomials related to (\ref{Harper}).

Now it is important to note [\ref{WZ}] that 
the following set of relations can be obtained from
(\ref{de}) by substituting $\Psi(z)=\prod_{k=1}^{Q-1}(z-z_k)$
and setting $z=z_m$, $m=1,\dots,Q-1$: 
\begin{equation}
\frac{z_m^2+q}{qz_m^2+1}=q^Q\prod_{k=1}^{Q-1}
\frac{qz_m-z_k}{z_m-qz_k},\qquad m=1,\dots,Q-1.\label{ba1}
\end{equation}
Collecting the coefficients at $z^{Q-1}$ in (\ref{de}) gives the
expression for the energy:
\begin{equation} 
\varepsilon=iq^Q(q-q^{-1})\sum_{m=1}^{Q-1}z_m.\label{e}
\end{equation}

Because of the analogy to one-dimensional integrable models,
where a similar set of equations is obtained, expressions
(\ref{ba1},\ref{e}) are called Bethe ansatz equations. 
The Bethe ansatz equations for integrable models are solved
in the limit  of infinite system. In our case this limit corresponds 
to $Q\to\infty$. Thus, we can hope 
to obtain from (\ref{ba1},\ref{e}) information about the spectrum at
irrational $\alpha$.

The analysis of (\ref{ba1},\ref{e}) is only in its beginnings so far.
The system (\ref{ba1},\ref{e}) was investigated numerically 
(in the case of $\varepsilon=0$ analytically) in [\ref{Hats}]. 
An analytical approach based on the string hypothesis is 
proposed in [\ref{ATW}]. 

In Section 2 we will obtain a solution (that is the distribution 
function of the roots $z_k$) in the limit as $\alpha=1/Q\to 0$, 
thus explaining some numerical results of [\ref{Hats}].
To obtain this solution, we will consider besides (\ref{ba1})
another set of Bethe ansatz equations which follow from (\ref{de}).

We find the solution under assumption that 
for $\alpha=1/Q$ the roots lie on the unit circle, and their
limit distribution function $\rho(\phi)$ depends piece-wise 
smoothly on the angle 
$\phi$.\footnote{As usual, $Q\rho(\phi)d\phi$ is the number of
roots in the interval $d\phi$.} The function $\rho(\phi)$ satisfies 
an integral equation that follows from the
Bethe ansatz equations. Besides $\rho(\phi)$, another important
function, $\chi(\phi)$,
naturally arises from the Bethe ansatz equations. It differs from unity
only when $\rho(\phi)={\rm const}=1/\pi$ and describes 
in this case the changing with $\phi$ exponentially small in $Q$
correction to the constant distance between neighboring roots. 
Note that a nonconstant
$\rho(\phi)$ implies the increase of order $1/Q$ in this distance 
with $\phi$.

In Section 3, using the expressions for $\rho(\phi)$ and $\chi(\phi)$,
we obtain the upper boundary
$\varepsilon_{\rm max}$ of the spectrum of (1) for small $\alpha$.
The result is $\varepsilon_{\rm max}=4-2\pi/Q+o(1/Q)$. The value $2\pi$
($-2\pi$) is the tangent of the lower (upper) boundary of the Hofstadter
butterfly for small magnetic field. Alternatively, this result 
can be obtained using semiclassical considerations [\ref{W}, \ref{B}]. 
Hopefully, the ideas introduced in the present work will be helpful in the 
cases where semiclassical techniques cannot be applied.

\section{Distribution of the roots of the Bethe ansatz equations for
$\alpha=1/Q\to 0$}
     
In order to obtain the results of this work, we will also need another set
of Bethe ansatz equations which is derived from (\ref{de}) by first replacing
$z$ with $qz$ and then substituting $z=z_m$, $m=1,\dots,Q-1$. We get
\begin{equation}
\frac{\varepsilon}{i(q^{-1}z_m^{-1}+q^2z_m)}=
\prod_{k=1}^{Q-1}\frac{q^2z_m-z_k}{qz_m-z_k},\qquad m=1,\dots,Q-1.
\label{ba2}
\end{equation}

Henceforth, we will be interested in the simplest case of the Harper equation
when $\alpha=1/Q$ ($P=1$, $\gamma=\pi/Q$ ) and $Q$ is large.
Numerical data [\ref{Hats}] suggests that 
if $\alpha=1/Q$ all the roots $z_k$ lie on the unit circle, they are simple,
and their distribution becomes continuous as $Q\to\infty$. 
Let us therefore assume $z_k=e^{i\phi_k}$, $\phi_k\in{\bf R}$, substitute this
expression into (\ref{ba1},\ref{e},\ref{ba2}), and then try to find from 
these equations a piece-wise smooth  
distribution function $\rho(\phi)$, $\phi\in[-\pi,\pi]$ in the
limit as $Q\to\infty$. After the substitution $z_k=e^{i\phi_k}$,
the equations (\ref{ba1},\ref{ba2}) take the form:
\begin{equation}
\frac{\cos(\phi_m-\gamma/2)}{\cos(\phi_m+\gamma/2)}=
-\prod_{k=1}^{Q-1}\frac{\sin{1\over2}(\phi_m-\phi_k+\gamma)}
{\sin{1\over2}(\phi_m-\phi_k-\gamma)},\label{bap1}
\end{equation}
\begin{equation}
\frac{-\varepsilon}{2\cos(\phi_m+3\gamma/2)}=
\prod_{k=1}^{Q-1}\frac{\sin{1\over2}(\phi_k-\phi_m-2\gamma)}
{\sin{1\over2}(\phi_k-\phi_m-\gamma)}.\label{bap2}
\end{equation}

\begin{figure}
\centerline{\psfig{file=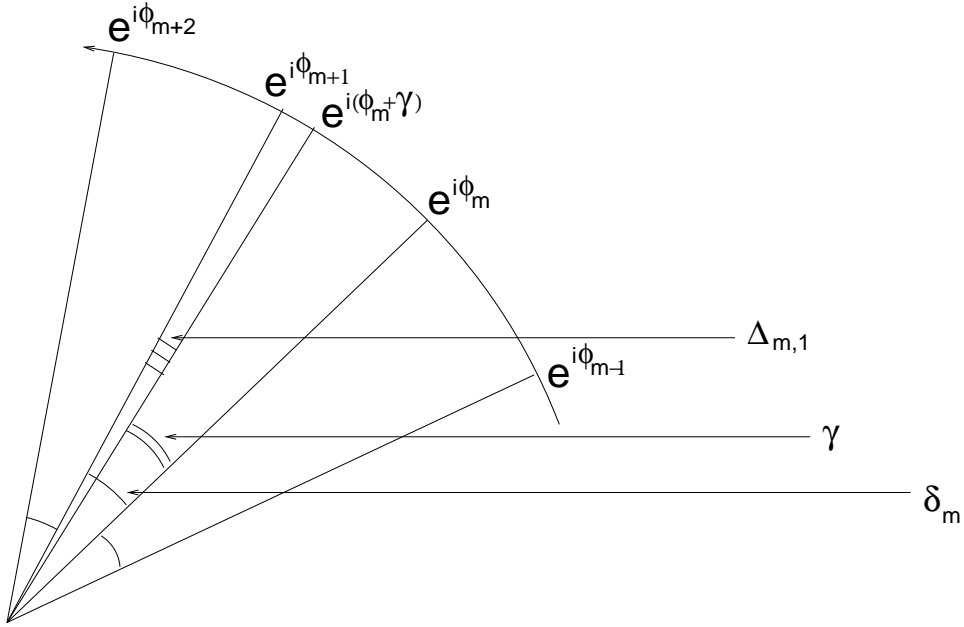,width=5in,angle=-90}}
\vspace{0.5cm}
\caption{A few consecutive roots $z_k=e^{i\phi_k}$ of the Bethe ansatz 
equations near the point $k=m$ are shown as they lie with respect to each
other for sufficiently large $Q$. The angles 
$\delta_m$, $\gamma$, $\Delta_{m,1}$ are, of course, increased.} 
\label{angles}
\end{figure}

We set $\phi_{k+1}-\phi_k=\delta_k$. According to our assumption
$\delta_k$ is of order $1/Q$ so that $\rho(\phi_k)=1/Q\delta_k$.
Smoothness of the distribution function implies that
$\phi_k-\phi_{k\pm p}=\mp p\delta_k$ for $p=1,\dots, N$
up to at most $O(N^2/Q^2)$
in the limit as $N,Q\to\infty$ in such a way that $N/Q\to 0$.
Let us consider equations (\ref{bap2}) in this limit. The product is separated
into three distinct parts. 
First, the contribution of the two possible points 
$\phi_{m+n}$ and  $\phi_{m+2n}$ such that $\phi_{m+n}-\phi_m-\gamma=o(1/Q)$
and $\phi_{m+2n}-\phi_m-2\gamma=o(1/Q)$,
where  $n$ is a certain integer which is the same
in the region of angles $\phi_k$ from $k=m-N$ to $k=m+N$. 
Define $\Delta_{m,i}=\phi_{m+i}-\phi_m-\gamma$ (See Figure \ref{angles}). Then 
$\phi_{m+2n}-\phi_m-2\gamma=\Delta_{m+n,n}+\Delta_{m,n}$. 
Note that if to the main order in $1/Q$ the distance
$\delta_m\ne\gamma,\gamma/2,\gamma/3,\dots$, then, whatever $i$, 
both $\Delta_{m,i}=O(1/Q)$ and $\Delta_{m+i,i}=\Delta_{m,i}+o(1/Q)$. 
In this case the ratio 
$\sin{1\over2}(\phi_{m+2n}-\phi_m-2\gamma)/
\sin{1\over2}(\phi_{m+n}-\phi_m-\gamma)$ is always equal to 2 in the limit
and, as we shall see, it is not
necessary to distinguish it as a separate contribution.
However, when $\delta_m=\gamma,\gamma/2,\gamma/3,\dots$ up to $o(1/Q)$, 
this ratio is an unknown and should be treated separately.
Below only the case $\delta_m=\gamma=\pi/Q$ will be relevant because
in the solution we obtain, $\delta_m\ge\pi/Q$ for any $\phi_m$.

The second contribution to the product comes from the rest of the points
$\phi_{m\pm p}$, where $p=1,\dots, N$. All the points not included into these
two groups give the third contribution. 

Thus, in the limit $N,Q\to\infty$, $N/Q\to 0$, which we denote just ``lim'',
we can write (\ref{bap2}) as follows
(denoting $\phi=\lim\phi_m$, $\phi\in[-\pi,\pi]$):
\begin{equation}
\frac{-\varepsilon}{2\cos\phi}=\lim ABC.\label{ABC}
\end{equation}
Here $A$ stands for the contribution of the points with the indices
$k=m+n$ and $k=m+2n$.
Expanding the sinus in $1/Q$, we can write:
\begin{equation}
\eqalign{
A=\frac{\sin{1\over2}(n\delta_m-2\gamma)}
{\sin{1\over2}(\phi_{m+n}-\phi_m-\gamma)}
\frac{\sin{1\over2}(\phi_{m+2n}-\phi_m-2\gamma)}
{\sin{1\over2}(2n\delta_m-\gamma)}\sim\\
\left(1+\frac{\Delta_{m+n,n}}{\Delta_{m,n}}\right)
\frac{n\delta_m-2\gamma}{2n\delta_m-\gamma}.}
\end{equation}
The factor $B$ denotes the contribution of the points with the indices
$k=m$, $k=m-n$, $k=m-2n$, and also $k=m\pm p$, $p=1,\dots,N$, $p\ne n,2n$:
\begin{equation}
\eqalign{
B=2 \frac{\sin{1\over2}(n\delta_m+2\gamma)}
{\sin{1\over2}(n\delta_m+\gamma)}
\frac{\sin{1\over2}(2n\delta_m+2\gamma)}
{\sin{1\over2}(2n\delta_m+\gamma)}
\prod_{p=1\atop p\ne n,2n}^N
\frac{\sin{1\over2}(p\delta_m-2\gamma)\sin{1\over2}(p\delta_m+2\gamma)}
{\sin{1\over2}(p\delta_m-\gamma)\sin{1\over2}(p\delta_m+\gamma)}\sim\\
2\frac{1+2\gamma/n\delta_m}{1+\gamma/2n\delta_m}
\prod_{p=1\atop p\ne n,2n}^N
\frac{1-(2\gamma/p\delta_m)^2}{1-(\gamma/p\delta_m)^2}.}
\end{equation}
Thus 
\begin{equation}
\eqalign{
AB\sim\left(1+\frac{\Delta_{m+2n-1}}{\Delta_{m+n-1}}\right)
\frac{1-(2\gamma/n\delta_m)^2}{1-(\gamma/2n\delta_m)^2}
\prod_{p=1\atop p\ne n,2n}^N
\frac{1-(2\gamma/p\delta_m)^2}{1-(\gamma/p\delta_m)^2}\sim\\
\left(1+\frac{\Delta_{m+2n-1}}{\Delta_{m+n-1}}\right)
\cos\left(\frac{\pi\gamma}{\delta_m}\right),}\label{AB}
\end{equation}
where we used the formula $\cos(\pi a/2)=\prod_{k=0}^\infty(1-a^2/(2k+1)^2)$.
Finally, $C$ in (\ref{ABC}) is the contribution of the points 
with the indices $k$ outside 
the range $m-N,\dots,m+N$. We denote the latter restriction by primes 
at the sums and products:
\begin{equation}
\eqalign{
C=\prod_k\mathop{^{'}}\frac{\sin{1\over2}(\phi_m-\phi_k+2\gamma)}
{\sin{1\over2}(\phi_m-\phi_k+\gamma)}\sim
\prod_k\mathop{^{'}}\frac{1+\gamma\cot{1\over2}(\phi_m-\phi_k)}
{1+{\gamma\over2}\cot{1\over2}(\phi_m-\phi_k)}\sim\\
\exp\left({\gamma\over2}\sum_k\mathop{^{'}}\cot{1\over2}(\phi_m-\phi_k)
\right),}\label{C}
\end{equation}
where the last product was replaced by the exponent of its logarithm.
The small-parameter expansion of the logarithm is valid here because 
in this range of $k$ the quantity
$\gamma\cot{1\over2}(\phi_m-\phi_k)$ is of order $1/N$ or less.  
Note that because we take the limit $N,Q\to\infty$ in such a way that 
$N/Q\to0$, the factor (\ref{C}) includes the contribution of all roots 
of the Bethe ansatz equations except for those lying in the 
infinitesimally small neighborhood of $z_m=e^{i\phi_m}$. Replacing the sum
in (\ref{C}) by the integral and inserting (\ref{AB}), (\ref{C}) into 
(\ref{ABC}), we obtain:
\begin{equation}
\eqalign{
\frac{-\varepsilon}{2\cos\phi}=
\left(1+\lim\frac{\Delta_{m+n,n}}{\Delta_{m,n}}\right)
\cos(\pi^2\rho(\phi))\times\\
\exp\left({\pi\over2}V.p.\int_{-\pi}^{\pi}\cot{1\over2}(\phi-\omega)
\rho(\omega)d\omega\right).}
\label{bapp2}
\end{equation}

A similar analysis of (\ref{bap1}) yields the result:
\begin{equation}
1=\lim\left[\frac{\Delta_{m,n}}{\Delta_{m-n,n}}\right]
\exp\left(\pi V.p.\int_{-\pi}^{\pi}\cot{1\over2}(\phi-\omega)
\rho(\omega)d\omega\right).
\label{bapp1}
\end{equation}
We see that the function 
$\chi(\phi)\equiv\lim(\Delta_{m,n}/\Delta_{m-n,n})$ is continuous.\footnote{
We omit the index $n$ of $\chi(\phi)$ because in the solution we will
obtain, it is always $n=1$.}
Hence, we can substitute
$\lim(\Delta_{m+n,n}/\Delta_{m,n})=\chi(\phi)$,
which is expressed from (\ref{bapp1}), in (\ref{bapp2}). Thus,
\begin{equation}
\frac{-\varepsilon}{4\cos\phi}=
\cosh\left\{
{\pi\over2}V.p.\int_{-\pi}^{\pi}\cot{1\over2}(\phi-\omega)
\rho(\omega)d\omega\right\}\cos(\pi^2\rho(\phi)).\label{ie}
\end{equation}

Substituting here the energy
\begin{equation}
\varepsilon=2\pi\int_{-\pi}^{\pi}e^{i\omega}\rho(\omega)d\omega,\label{ep}
\end{equation}
we obtain an integral equation for the distribution function $\rho(\phi)$.
It is not the only possible form of the integral equation for $\rho(\phi)$
we can obtain. We can derive others by considering other sets of Bethe ansatz
equations which follow from (\ref{de}) in a similar manner as (\ref{ba2}).

Equation (\ref{ie}) reflects the well-known symmetry of the roots
$\{z_m\}$. They lie symmetrically with respect to the real axis.
So $\rho(-\phi)=\rho(\phi)$.

Note that the function $\rho(\phi)$ must also satisfy the normalization 
condition 
\begin{equation}
\int_{-\pi}^{\pi}\rho(\phi)d\phi=1.\label{norm}
\end{equation}

Eliminating the exponents from (\ref{bapp2}) and
(\ref{bapp1}) we get a simple relation between $\chi(\phi)$ and
$\rho(\phi)$:
\begin{equation}
-\varepsilon\sqrt{\chi(\phi)}=2\cos\phi\cos(\pi^2\rho(\phi))
(1+\chi(\phi)).\label{rhochi}
\end{equation}

We can guess the form of the solution $\rho(\phi)$ 
using the following argument. 
Since $\rho(\phi)$ is even, it follows from (\ref{bapp1}) 
that $\chi(0)=\chi(\pi)=1$. Thus, in some neighborhood
of the points $\phi=0,\pi$ equation
(\ref{rhochi}) simplifies to
the form $\varepsilon=-4\cos\phi \cos(\pi^2\rho(\phi))$.\footnote{
Comparing this with (\ref{bapp2}) we see 
that only the points lying in the infinitesimally small
neighborhood of $z_m$, whose contribution is described by (\ref{AB}), 
influence $\rho(\phi_m)$.}
This can hold only in the range $|\cos\phi|\ge\varepsilon/4$. 
Outside of it, $\chi(\phi)$ can no longer be 1, which means that 
$\rho(\phi)$ can be only a piece-wise constant function there taking 
some of the values
$1/\pi,2/\pi,...$ or zero. A precise form of $\rho(\phi)$ is given by the
following
\bigskip

{\bf Statement}.
$\rho(\phi)={1\over\pi^2} \arccos\frac{\varepsilon}{-4\cos\phi}$
if $\phi\in[-\psi,\psi]\cup [\pi-\psi,\pi]\cup [-\pi,-\pi+\psi]$,
where $\psi=\arccos\frac{\varepsilon}4$, $0\le\arccos(x)\le\pi$;
$\varepsilon\in[0,4]$.

$\rho(\phi)=1/\pi$ if $\phi\in[\psi,\pi/2)\cup (-\pi/2,-\psi]$; 
  
\begin{figure}
\centerline{\psfig{file=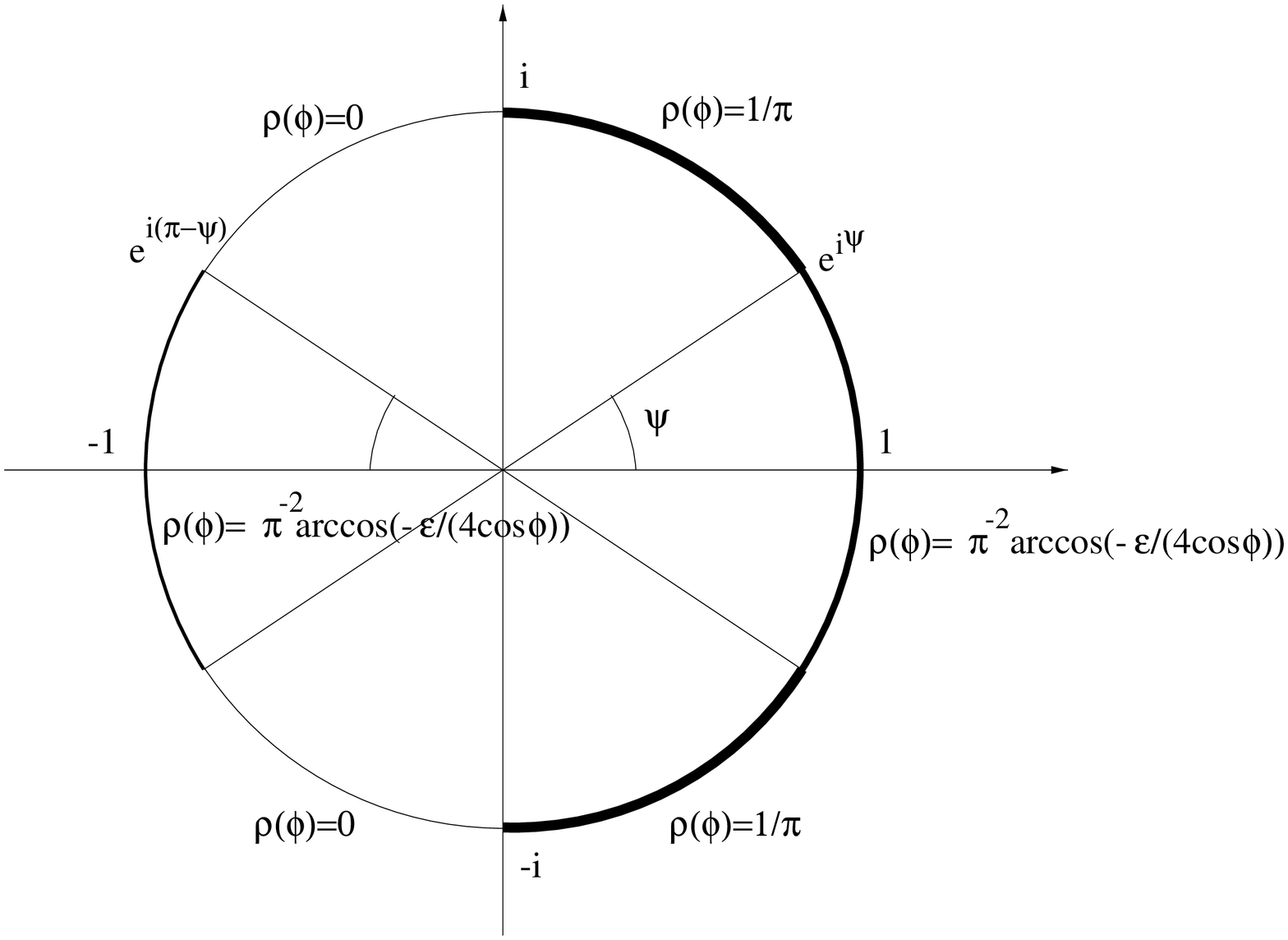,width=5in,angle=-90}}
\centerline{\psfig{file=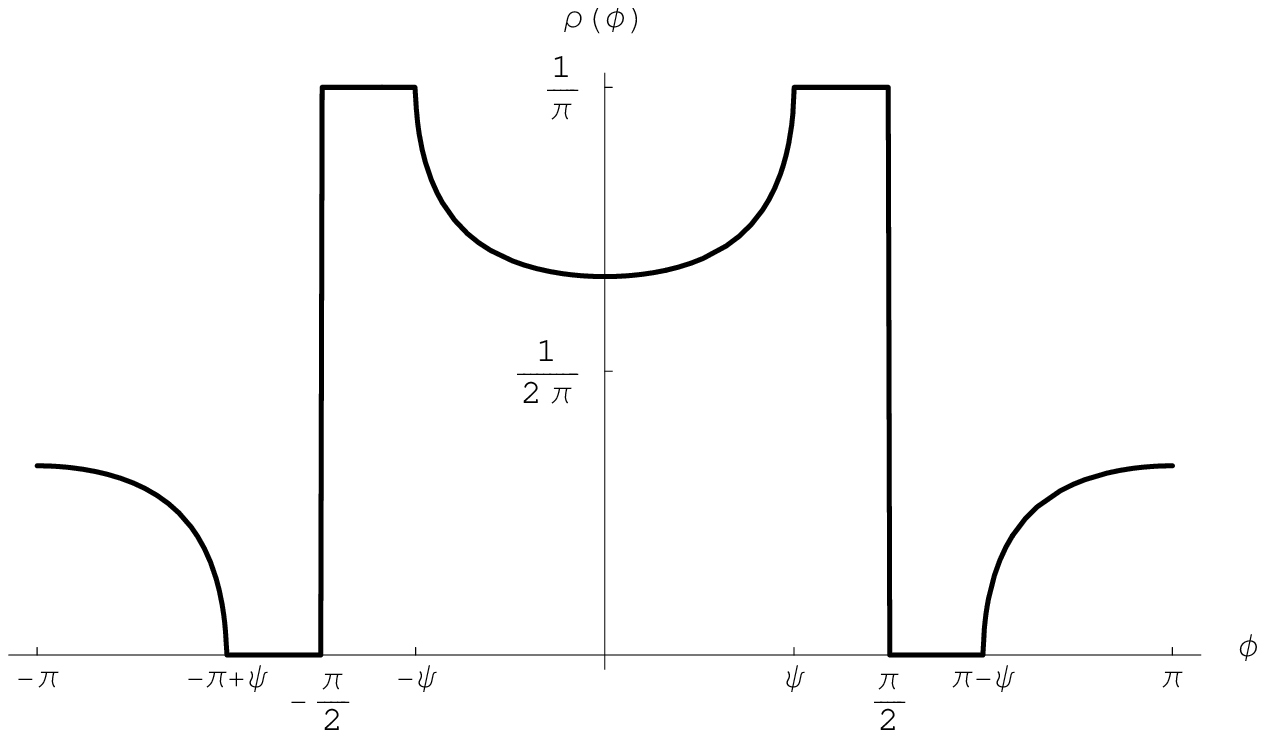,width=6.3in,angle=0}}

\caption{Distribution of the roots of the Bethe ansatz equations
in the limit as $\alpha=1/Q\to 0$. In the upper figure, the regions on the
unit circle are shown where the roots lie with the density
$\rho(\phi)$, where $\phi$ is the angle. In the lower figure, 
$\rho(\phi)$ for some $\varepsilon$ is plotted as a function of the angle.}
\label{distr}
\end{figure}

$\rho(\phi)=0$ if $\phi\in[\pi/2,\pi-\psi]\cup [-\pi+\psi,-\pi/2]$.
\bigskip
(See Figure \ref{distr}.)

The same but reflected with respect to 
the imaginary axis distribution corresponds to the interval
$\varepsilon\in[-4,0]$.

Henceforth, we always assume $\varepsilon\in[0,4]$.

We verify the statement substituting this expression for $\rho(\phi)$
in (\ref{ie}), (\ref{ep}), (\ref{norm})  and integrating.
For (\ref{ep}) we have:
\begin{equation}
\eqalign{
2\pi\int_{-\pi}^{\pi}e^{i\omega}\rho(\omega)d\omega=\\
2\left(\left\{\int_\psi^{\pi/2}+\int^{-\psi}_{-\pi/2}\right\}e^{i\omega}
d\omega+
\left\{\int_{-\psi}^{\psi}+\int^{\pi+\psi}_{\pi-\psi}\right\}
{1\over\pi}e^{i\omega}\arccos\frac{\varepsilon}{-4\cos\omega}d\omega\right)=\\
4-{8\over\pi}\int_0^\psi \cos\omega\arccos\frac{\varepsilon}{4\cos\omega}
d\omega=\\
4-\frac{2\varepsilon}{\pi}\int_0^\nu\arctan x\frac{xdx}{\sqrt{\nu^2-x^2}}=
4-\frac{2\varepsilon\nu}{\pi}\int_{\nu^{-1}}^\infty
\frac{\sqrt{u^2-\nu^{-2}}du}{u(1+u^2)}=\varepsilon,}
\end{equation}
where  $\nu=\sqrt{(4/\varepsilon)^2-1}$.
Henceforth, we take $\sqrt{x}>0$ if $x>0$. 

\begin{figure}
\centerline{\psfig{file=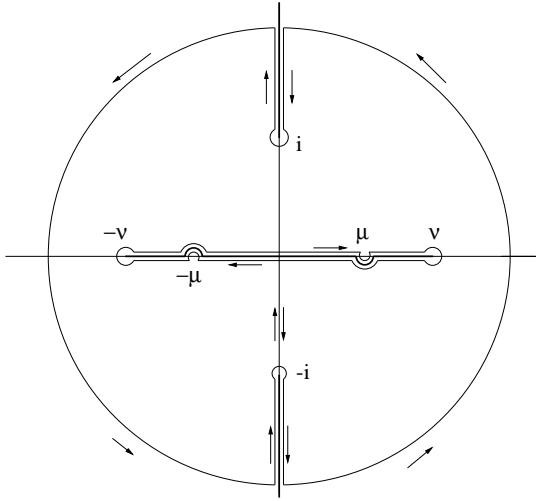,width=2.8in,angle=-90}}
\vspace{0.5cm}
\caption{The contour in the complex plane used to evaluate the integral
(\protect\ref{vpint}). The thick lines denote cuts. There are two simple poles
at the points $\mu$ and $-\mu$.}
\label{contour}
\end{figure}

Since $\rho(\phi)$ is an even function,
it is sufficient to verify expression (\ref{ie}) only for 
$\phi\in[0,\pi]$. 
There are three distinct cases: $\phi\in[0,\psi]$,
$\phi\in[\psi,\pi/2]$, $\phi\in[\pi-\psi,\pi]$.
Most of the relevant integrals are easy to evaluate.
Perhaps only the calculation of the following one 
is somewhat less straightforward:
\begin{equation}
\eqalign{
V.p. {1\over 2}\int_{-\psi}^\psi
\left\{\cot{1\over 2}(\phi-\omega)+\tan{1\over 2}(\phi-\omega)\right\}
\arccos\frac{\varepsilon}{4\cos\omega}d\omega=\\
V.p.\int_{-\psi}^\psi\frac{\arccos\{\varepsilon/(4\cos\omega)\}}
{\sin(\phi-\omega)}d\omega=
\frac{8\sin\phi}{\varepsilon}
V.p.\int_0^\nu\frac{x\arctan x\; dx}{(x^2-\mu^2)\sqrt{\nu^2-x^2}}=\\
\frac\pi{2}\ln\frac{1+\sin\phi}{1-\sin\phi},\label{vpint}}
\end{equation}
where $\nu=\sqrt{(4/\varepsilon)^2-1}$, 
$\mu=\sqrt{(4\cos\phi/\varepsilon)^2-1}$, $\cos\phi>\varepsilon/4$. 
The contour used for evaluation of the last integral in (\ref{vpint}) is
shown in Figure \ref{contour}.

From (\ref{bapp1}) we obtain the expression for $\chi(\phi)$ on the set
$\phi\in[\psi,\pi/2)\cup (-\pi/2,-\psi]$ 
where $\rho(\phi)=1/\pi$ 
(note that $\chi(\phi)=1$
if $\phi\in[-\psi,\psi]\cup [\pi-\psi,\pi]\cup [-\pi,-\pi+\psi]$):
\begin{equation}
\chi(\phi)=\lim\frac{\Delta_{m,1}}{\Delta_{m-1,1}}=
\frac{1-\sqrt{1-({4\over\varepsilon}\cos\phi)^2}}
{1+\sqrt{1-({4\over\varepsilon}\cos\phi)^2}},
\qquad\phi\in[\psi,\pi/2).\label{chi}
\end{equation}
Thus, in the region $\phi\in(\psi,\pi/2)$ we have
$\Delta_{m-1+k,1}=\chi(\phi)^{(k/Q)Q}\Delta_{m-1,1}$, $0<\chi(\phi)<1$,
which means that as $\phi$ increases from $\psi$, the distance between
the roots $\phi_m$ approaches the constant value $\pi/Q$ exponentially fast.
We shall use the function $\chi(\phi)$ in the next section to determine
the finite $Q$ correction to the boundary $\varepsilon=4$ of the spectrum. 

Because of the mentioned symmetry of the Bethe ansatz equations,
it is sufficient to write the expressions for
$\chi(\phi)$ only for $\phi\in[0,\pi]$. This symmetry implies
$\chi(-\phi)=\chi(\phi)^{-1}$.

\section{The boundaries of the spectrum of the Harper equation for small
$\alpha$.}
Consider now the simplest case $\varepsilon=4$ in more detail. 
In this case $\rho(\phi)=1/\pi$ if $\phi\in(-\pi/2,\pi/2)$ and zero otherwise.
Equation (\ref{chi}) takes the form
\begin{equation}
\chi(\phi)=\frac{1-\sin\phi}{1+\sin\phi},\qquad
\phi\in[0,\pi/2).\label{chi4}
\end{equation}
We shall calculate the largest finite $Q$ correction
to the energy $\varepsilon=4$.
It turns out that the shift (from the uniform distribution defined below) 
of the roots of the Bethe ansatz equations described
by (\ref{chi4}) is responsible for this correction. 
Since according to (\ref{chi4}),
$\Delta_{m,1}$ decreases exponentially as $\phi_m$ moves away from the point 
$\phi=0$, only an infinitesimally small neighborhood of  $\phi=0$ 
gives considerable contribution to the shift. Let us  
assume the following distribution of the roots $\phi_0=0$, 
$\phi_m=m\pi/Q+\sum_{i=0}^{m-1}\Delta_{i,1}$, 
$m=1,\dots,Q/2-1$, $\phi_{-m}=-\phi_m$ (we set $Q$ even for definiteness).  
The mentioned property of $\Delta_{i,1}$ implies that if $\phi_m\to\phi\ne 0$ 
as $Q\to\infty$, then
with an exponential accuracy
$\phi_m=m\pi/Q+(m/|m|)s/Q^\delta$, where 
$s/Q^\delta=\sum_{i=0}^\infty\Delta_{i,1}$.
The quantities 
$s$ and $\delta$ are to be found from the Bethe ansatz equations.
Henceforth, we assume $m>0$.
Further, set
\begin{equation}
\frac{\sin(\Delta_{m,1}/2)}{\sin(\Delta_{m-1,1}/2)}=
\chi\left(\frac{\pi m}Q\right)
\left\{1+\frac{r_m}{Q^\beta}+o\left({1\over Q^\beta}\right)\right\}.
\end{equation}
Retaining now in equations
(\ref{bap1}) not only the largest in $1/Q$ quantities (of order $1$)
as before, but also the next largest 
(that is of the order of $1/Q$,  $1/Q^\beta$,
$1/Q^\delta$), we obtain with this accuracy the following 
expression for the logarithm of (\ref{bap1})
\begin{equation}
\eqalign{
{\pi\over Q}\tan{\pi m\over Q}=\ln\chi(\frac{\pi m}Q)+{r_m\over Q^\beta}-\\
{2s\over Q^\delta}\left\{{\pi\over 2Q}\sum_{k=-(Q/2-1)}^0
\sin^{-2}\frac{\pi(m-k)}{2Q}
\right\}+{\pi\over Q}\sum_k{^{'}}\cot\frac{\pi(m-k)}{2Q}.}\label{med}
\end{equation}
Replacing the first sum over $k$ in (\ref{med}) by the integral gives
\begin{equation}
{\pi\over 2Q}\sum_{k=-(Q/2-1)}^0\sin^{-2}\frac{\pi(m-k)}{2Q}\sim
\int_{-\pi/2}^0\sin^{-2}\frac{{\pi m\over Q}-x}2\; {dx\over 2}=
\frac 2{1-\cos{\pi m\over Q}+\sin{\pi m\over Q}}.
\end{equation}
As to the second sum over $k$ in (\ref{med}), we need to evaluate it 
up to $1/Q$. Using the Poisson summation formula, we have
\begin{equation}
\eqalign{
{\pi\over Q}\sum_k{^{'}}\cot\frac{\pi(m-k)}{2Q}=
{\pi\over Q}\sum_{k=-(Q/2-1)}^{-Q/2+2m}\cot\frac{\pi(m-k)}{2Q}=\\
{\pi\over Q}
\int_{-{\pi\over 2}+{\pi\over 2Q}}
^{-{\pi\over 2}+2{\pi m\over Q}+{\pi\over 2Q}}
\cot\frac{{\pi m\over Q}-x}2
\sum_{k=-\infty}^\infty\delta\left(x-{\pi k\over Q}\right)\;dx=\\
\int_{-{\pi\over 2}+{\pi\over 2Q}}
^{-{\pi\over 2}+2{\pi m\over Q}+{\pi\over 2Q}}
\cot\frac{{\pi m\over Q}-x}2\;dx+
2\sum_{k=1}^\infty
\int_{-{\pi\over 2}+{\pi\over 2Q}}
^{-{\pi\over 2}+2{\pi m\over Q}+{\pi\over 2Q}}
\cot\frac{{\pi m\over Q}-x}2
\cos(2Qkx)\; dx=\\
\ln\frac{1+\sin{\pi m\over Q}}{1-\sin{\pi m\over Q}}+
{\pi\over Q}\tan{\pi m\over Q}+O(1/Q^2).}
\end{equation}
Here the value of the last sum over $k$ is estimated by
integration by parts. It is of order $1/Q^2$.
Thus, (\ref{med}) reduces to
\begin{equation}
\frac{r_m}{Q^\beta}-\frac{4s/Q^\delta}{1-\cos{\pi m\over Q}+
\sin{\pi m\over Q}}=0.\label{Q1}
\end{equation}
A similar analysis of (\ref{bap2}) gives the result
\begin{equation}
-\frac{\pi}{2Q\cos{\pi m\over Q}}=\frac{s}{Q^\delta}\left(
1-\tan{\pi m\over Q}-\frac{2}{1-\cos{\pi m\over Q}+\sin{\pi m\over Q}}\right)+
\frac{r_m\cos^2{\pi m\over Q}}{2Q^\beta(1+\sin{\pi m\over Q})}.\label{Q2}
\end{equation}
From (\ref{Q1},\ref{Q2}) we get $\delta=\beta=1$, $s=\pi/2$. 
The equation for the energy reads
$\varepsilon=4(1-s/Q)+o(1/Q)=4-2\pi/Q+o(1/Q)$.
(Correspondingly, the lower boundary of the spectrum is 
$\varepsilon=-4+2\pi/Q+o(1/Q)$.)
The same correction $2\pi/Q$ can be obtained from  
semiclassical considerations [\ref{B}].

\section{Conclusions}
In this work we have studied the distribution of the roots $z_k$ of the 
Bethe ansatz
equations for the Harper equation in the limit as $Q\to\infty$ while
the commensurability parameter $\alpha=P/Q$, $P=1$.
We considered two related sets of the Bethe ansatz equations 
to the main order as $1/Q\to 0$ and obtained functions $\rho(\phi)$ and 
$\chi(\phi)$ that characterize the distribution. It is a continuous
distribution of the roots on the unit circle. 
 
Considering the Bethe ansatz equations to the next (smaller) order in $1/Q$
we calculated the small $\alpha$
correction to the upper and lower boundaries of the spectrum of the Harper
equation. 

One more step further could be to extend our treatment to the case
of any fixed $P$, while $Q\to\infty$. Numerical data suggests that in 
this case the roots $z_k$ for the largest eigenvalue $\varepsilon_{\rm max}$ 
become continuously distributed on $P$ smooth curves.

\section{Acknowledgements}
I am grateful to J. Bellissard, H. Castella, and 
S. Mikhailov for useful discussions during preparation of
this work.

\section{Appendix}
Here we derive the $q$-difference equation (\ref{de}).

\begin{figure}
\centerline{\psfig{file=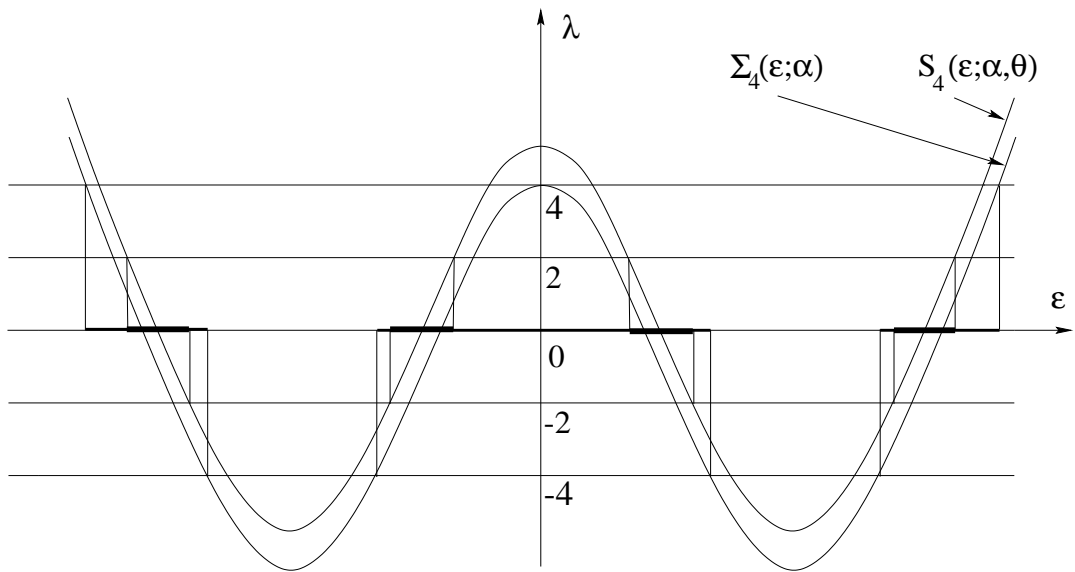,width=5.6in,angle=-90}}
\vspace{0.5cm}
\caption{The plot of $S_Q(\varepsilon;\alpha,\theta)$ 
and $\Sigma_Q(\varepsilon;\alpha)$ for $Q=4$ and some fixed
$\theta$. (For $Q=4$ the curves do not depend on $P$: $P=1$ or $P=3$.)
The thickest lines on the axis $\varepsilon$ are the intervals of the spectrum 
$\sigma_M(\alpha,\theta)$ of $M$.
The less thick lines are the intervals of the spectrum $\sigma_H(\alpha)$.
Two of the latter intervals touch at $\varepsilon=0$.
The curves shown are not the actual ones but qualitatively similar.}
\label{figapp}
\end{figure}

Assume $\alpha=P/Q$ with coprime integers $P,Q$. Then
(1) is the eigenvalue equation for a periodic tridiagonal matrix (denote this
matrix $M$) with the period $Q$ (that is the matrix elements 
$M_{ik}=M_{i+Q\,k+Q}$). The approach to the spectral problem for such a 
matrix is well known [e.g., \ref{Toda}]. We look for solutions to
(\ref{Harper}) in the
form $\psi_{j+Qk}=e^{i\omega k}\mu_j$, $j=0,1,\dots,Q-1$, where $\omega$ is an
arbitrary real number. This reduces (\ref{Harper}) to the
following eigenvalue equation
\begin{equation}
\eqalign{
\mu_{n-1}+(q^{2n}e^{i\theta}+q^{-2n}e^{-i\theta})\mu_n+\mu_{n+1}=
\varepsilon\mu_n,\\
n=0,1,\dots,Q-1;\qquad q=e^{i\pi\alpha};\qquad
\mu_{-1}=e^{-i\omega}\mu_{Q-1};\quad\mu_{Q}=e^{i\omega}\mu_{0}.}\label{2}
\end{equation}
for a $Q\times Q$ matrix (denote it $L$). It is easy to show that 
$\det(L-\varepsilon I)=(-1)^Q(S_Q(\varepsilon;\alpha,\theta)-2\cos\omega)$,
where $S_Q(\varepsilon;\alpha,\theta)$ is a polynomial of degree $Q$
in $\varepsilon$, which is called the
discriminant of $M$. The spectrum of $M$ is obviously given by the equation
$S_Q(\varepsilon;\alpha,\theta)=2\cos\omega$, $\omega\in[0,\pi]$. As is known,
the spectrum  consists of
{\it exactly} $Q$ intervals, the image of $[-2,2]$ under the inverse of the
transform  $S_Q(\varepsilon;\alpha,\theta)=\lambda$ 
(see Figure \ref{figapp}).\footnote{One can prove this by showing that 
$S_Q(M;\alpha,\theta)=E$, where the matrix elements of $E$ are
$E_{ij}=1$ if $|i-j|=Q$ and zero otherwise. The spectrum of $E$ is 
obviously $2Q$-degenerate, while the spectrum of the tridiagonal matrix $M$
can be at most doubly degenerate.}
Hence, all the extremal points of $S_Q(\varepsilon;\alpha,\theta)$ are
maxima and minima; moreover, at all maxima points
$S_Q(x_{\rm max};\alpha,\theta)\ge2$ and at all minima points
$S_Q(x_{\rm min};\alpha,\theta)\le-2$.

As we mentioned in the introduction,
the spectrum $\sigma_H(\alpha)$ of the Azbel-Hofstadter problem is the union
over all real $\theta$ of the spectra of $M$. The Chambers formula
[\ref{Chamb}] says that
the polynomial $\Sigma_Q(\varepsilon;\alpha)=
S_Q(\varepsilon;\alpha,\theta)+2\cos\theta Q$ does not depend on $\theta$,
which implies that 
$\sigma_H(\alpha)$ is the image of
$[-4,4]$ under the inverse of the transform 
$\Sigma_Q(\varepsilon;\alpha)=\lambda$. 
Considering the Chambers formula at $\theta=0$ and 
$\theta=\pi/Q$, we see that 
$\Sigma_Q(x_{\rm max};\alpha)\ge 4$ and
$\Sigma_Q(x_{\rm min};\alpha)\le -4$ at all its extremal points. 
So that $\sigma_H(\alpha)$ also consists of $Q$ intervals
(see Figure \ref{figapp}).

Let us now turn to the derivation of (\ref{de}).
Set $\omega=0$.
Substituting $\mu_n=\sum_{k=0}^{Q-1}U_{nk}e^{ik\theta}\phi_k$, where
$U_{nk}=q^{2nk}$, $n,k=0,1,\dots,Q-1$,\footnote{The matrix $U$ is
nondegenerate. $\det U\ne 0$ because $UU^*=QI$.}
in (\ref{2}), we have
\begin{equation}
\eqalign{
\phi_{k-1}+(q^{2k}+q^{-2k})\phi_k+\phi_{k+1}=
\varepsilon\phi_k,\\
k=0,1,\dots,Q-1;\qquad \phi_{-1}=e^{iQ\theta}\phi_{Q-1};\quad
\phi_{Q}=e^{-iQ\theta}\phi_{0}.}\label{3}
\end{equation}

By the nondegenerate transformation
\begin{equation}
\phi_k=\sum_{n=0}^{Q-1}q^{(k+1/2)^2+2kn} \zeta_n\label{4}
\end{equation}
we get rid of the middle term in (\ref{3}):
\begin{equation}
\eqalign{
(1+q^{2n})\zeta_{n-1}+(1+q^{-2(n+1)})\zeta_{n+1}=
\varepsilon\zeta_n,\\
n=0,1,\dots,Q-1;\qquad \zeta_{-1}=\zeta_{Q-1};\quad
\zeta_{Q}=\zeta_{0}.}\label{5}
\end{equation}
In order to obtain (\ref{5}) we had to assume the following condition:
for $Q$ even $Q\theta=\pi (\mod\; 2\pi)$; for $Q$ odd $Q\theta=0 
(\mod\; 2\pi)$.
If $Q$ is even, we can make the shift $\zeta_{n+Q/2}=\xi_n$, so that 
we get the equations
\begin{equation}
\eqalign{
(1-q^{2n})\xi_{n-1}+(1-q^{-2(n+1)})\xi_{n+1}=
\varepsilon\xi_n,\\
n=0,1,\dots,Q-1}\label{6}
\end{equation}
with zero boundary conditions ($\xi_{-1}$ and $\xi_Q$ may be arbitrary,
e.g., zero).
Further transformation $\xi_n=p_n i^n q^{n(n+1)/2}$ gives
\begin{equation}
\eqalign{
i(q^n-q^{-n})p_{n-1}+i(q^{n+1}-q^{-(n+1)})p_{n+1}=
\varepsilon p_n,\\
n=0,1,\dots,Q-1.}\label{7}
\end{equation}

Now multiplying (\ref{7}) by $z^n$ and then taking the sum over $n$ from 
$n=0$ to $Q-1$, we get
\begin{equation}
i(z^{-1}+qz)\Psi(qz)-i(z^{-1}+q^{-1}z)\Psi(q^{-1}z)=\varepsilon\Psi(z),
\label{de2}
\end{equation}
which is equation (\ref{de}). Here
$\Psi(z)=\sum_{n=0}^{Q-1}z^np_n$.
We derived (\ref{de2}) on condition that $\omega=0$, 
$Q\theta=\pi (\mod\; 2\pi)$, and $Q$ is even (hence, $P$ is odd).
It is easy to derive the same equation (\ref{de2}) for the two other possible
cases (that is $P$ even, $Q$ odd, and $P$ odd, $Q$ odd) using slightly 
different from (\ref{4}) transformations.
In all cases we have the same restriction  $\omega=0$, 
$Q\theta=\pi (\mod\; 2\pi)$. This implies that the solutions $\varepsilon$ of  
(\ref{de}) are the points satisfying 
$\Sigma_Q(\varepsilon;\alpha)=0$, so they lie inside the bands of
the spectrum $\sigma_H(\alpha)$ of the Azbel-Hofstadter problem.
As follows from the Chambers formula, they also lie in the bands of 
$\sigma_M(\alpha,\theta)$ for any $\theta$. Take any band of 
$\sigma_M(\alpha,\theta)$. It has a single point in it which is a solution of
(\ref{de}). As $\theta$ changes from $\theta=0$ to $\theta=\pi/Q$ the
band moves through this point from one boundary where  
$S_Q(\varepsilon;\alpha,0)=-2$ to the other where
$S_Q(\varepsilon;\alpha,\pi/Q)=2$.

\end{document}